\documentclass[prl,aps,floatfix,twocolumn,preprintnumbers]{revtex4}
\usepackage{epsfig,amsmath,amssymb,graphics,color,calc,times,wasysym}

\begin{document}

\title[]{Rotational averaging-out gravitational sedimentation of colloidal dispersions and phenomena}

\author{Djamel El Masri,$^{1}$\footnote{J.H.M.alMasri@uu.nl} Teun Vissers$^1$, Stephane Badaire$^1$, Johan C.P. Stiefelhagen$^1$, Hanumantha Rao Vutukuri$^1$, Peter Helfferich$^1$, Tian Hui Zhang$^2$, Willem K. Kegel$^2$, Arnout Imhof$^1$, and Alfons van Blaaderen$^{1}$\footnote{A.vanBlaaderen@uu.nl}}
\affiliation{$^{1}$Soft Condensed Matter, Debye Institute for Nanomaterials Science, Utrecht University, Princetonplein 5, 3584 CC Utrecht, The Netherlands. \\
$^{2}$Van't Hoff Laboratory for Physical and Colloid Chemistry, Utrecht University, Padualaan 8, 3584 CH Utrecht, The Netherlands.}

\date{\today}

\begin{abstract}
We report on the differences between colloidal systems left to evolve in the earth's gravitational field and the same systems for which a slow continuous rotation averaged out the effects of particle sedimentation on a distance scale small compared to the particle size. Several systems of micron-sized colloidal particles were studied: a hard sphere fluid, colloids interacting via long-range electrostatic repulsions above the freezing volume fraction, an oppositely charged colloidal system close to either gelation and/or crystallization, colloids with a competing short-range depletion attraction and a long-range electrostatic repulsion, colloidal dipolar chains, and colloidal gold platelets under conditions where they formed stacks. Important differences in the structure formation were observed between the experiments where the particles were allowed to sediment and those where sedimentation was averaged out. For instance, in the case of colloids interacting via long-range electrostatic repulsions, an unusual sequence of dilute-Fluid/dilute-Crystal/dense-Fluid/dense-Crystal phases was observed throughout the suspension under the effect of gravity, related to the volume fraction dependence of the colloidal interactions, whereas the system stayed homogeneously crystallized with rotation. For the oppositely charged colloids, a gel-like structure was found to collapse under the influence of gravity with a few crystalline layers grown on top of the sediment, whereas when the colloidal sedimentation was averaged out, the gel completely transformed into crystallites that were oriented randomly throughout the sample. Rotational averaging out gravitational sedimentation is an effective and cheap way to estimate the importance of gravity for colloidal self-assembly processes.
\end{abstract}

\maketitle

\section{Introduction}
Colloidal dispersions are often used as model systems to investigate a variety of phenomena in (soft) condensed matter physics, including crystallization, clustering/gelation, and the glass transition.\cite{Russel92,PuseyHouches89} Particles can interact with each other via e.g. excluded volume only,\cite{Pusey86} electrostatic interactions and van der Waals attractions.\cite{Russel92} However, external fields such as gravity and electric and magnetic fields can also influence the phase behavior either directly by changing the interparticle potentials, or indirectly by coupling to the osmotic pressure, or both. Due to their mesoscopic sizes, colloids have been used to study a variety of problems related to many particle statistical mechanics in- and out of equilibrium. It is experimentally possible to use relatively simple and powerful techniques, such as optical and confocal microscopy \cite{KegelSci00, WeeksSci00} or static and dynamic light scattering \cite{BernePecora, BrambillaPRL09, DjamelJSTAT09} to fully characterize their structural and dynamical properties. However, because colloids are much larger than atoms ($10^3-10^4$ times larger), and the interparticle cohesion is therefore relatively weak, gravity often plays a significant role and affects both the colloidal equilibrium phase behavior, as well as dynamical processes. On the single-particle level the gravitational length, $l_g=k_B T/G$, where $k_B T$ is the thermal energy and $G=\pi \sigma^3 \Delta \rho g /6 $ is the buoyant colloid weight (with $\sigma$ the particle diameter, $\Delta \rho= | \rho_{p} - \rho_{s} |$ the density mismatch between the particle and the solvent, and $g$ the gravitational acceleration), characterizes at what heights gravity will start to play an important role in the equilibrium phase behavior. If the gravitational length is sufficiently larger than the colloidal size the equation of state can be deduced from the density profile in a single sample. For this, the density profile should be measured as a function of height.\cite{RutgersPRL96, PiazzaPRL93, BeckhamJCP07, RoyallPRL07} Recent work has shown that this can be obtained quite effectively from real-space microscopy techniques.\cite{RoyallJPCM05} However, care has to be taken in the case of systems with long-range electrostatic interactions between the colloids, as the effect of the double layer complicates this scheme.\cite{RasaNature04} The P\'{e}clet number, $Pe$, the ratio between the effect of sedimentation against that of Brownian diffusion, is another dimensionless quantity that characterizes the influence of gravity, here with its effects on the particle dynamics:
\begin{equation}
Pe = \tau_{\mathrm B}/\tau_{\mathrm S} = \sigma/2 l_{\mathrm g} = \pi \sigma^4 \Delta \rho g / 12 k_{\mathrm B} T,
\label{Peclet}
\end{equation} 
where $\tau_{\mathrm B}$ and $\tau_{\mathrm S}$ are the time scales for Brownian diffusion and sedimentation respectively. When $Pe \ll 1$, gravity may be neglected, while in the case $Pe \gg 1$ the behavior of a colloidal dispersion will be strongly affected by gravity. From this relation (Eq. \ref{Peclet}), one can appreciate the strong effects of particle size and density on the P\'{e}clet number as this varies with the fourth power of the particle diameter, $Pe\varpropto\sigma^4$, and linearly with the density mismatch, $Pe\varpropto \Delta \rho$. For example, for a colloidal particle of 1 $\mu$m diameter and a density mismatch of $\Delta \rho = 0.10$ g/ml ($l_{\mathrm g} \simeq 10~\mu$m), the drift of the particle due to gravity is 10 times larger than that of Brownian diffusion. By choosing the right particle size and matching the density of the particles to that of the solvent, one can significantly reduce the influence of gravity in laboratory experiments. However, it is experimentally difficult to reduce the density difference $\Delta \rho$ significantly below $10^{-3}$ g/ml without extreme temperature control (better than 0.01 $^{\circ}$C) required to reach the condition of $\mu$-gravity.\cite{Kegel-Langmuir99, PeterLuNature08} In addition, the gravitational length and P\'{e}clet number are single particle properties; clearly, interactions at increased volume fractions can significantly modify the effects of gravity. For instance,  when colloids form solid-like structures, such as aggregates or crystallites, the relevant buoyant force is not that of a single particle, but rather that experienced by solid structures composed of many colloids. Because of the already mentioned strong size dependence of both the gravitational length and the sedimentation velocity, these collective effects are enormous and can already make the effects of gravity significant for nano-particle dispersions if they solidify. Moreover, a sedimenting piece of a colloidal solid, like a crystal or a gel, experiences significant shearing forces of the liquid during its fall, and sets up a convective counter flow as well. It was only recently realized that such flows can significantly alter the shape of the colloidal crystals formed.\cite{Zhu-Nature97} These effects alter the individual particle trajectories and consequently affect the growth mechanism, nucleation process and collective dynamics. Although less severe than in the case of solidification, in the case of attractive interactions in general particles will on average spend more time in a ``clustered state'' than in case of repulsive systems, thus increasing the effects of gravity. When the attractions are large enough to cause a gas-liquid phase separation, again the newly formed denser phase will quickly start to sediment. Therefore, it is important to gauge the effects of gravity, and to determine when experiments can be considered to be in the limit of \textit{zero}-gravity, especially for colloidal systems with a gravitational length smaller than the container size.  

Experiments which compare those performed at $1~g$ with the same experiments done in a microgravity environment are ultimately required to quantify the gravity related phenomena and to investigate the influence of weak forces on physico-chemical processes, such as phase separation and aggregation/gelation. This question is what drove scientists to perform experiments under micro-gravity conditions in so-called parabolic flights that can be performed by airplanes\cite{OkuboCPS2000,TsuchidaCPS2000}, and also in the Space Shuttle\cite{Zhu-Nature97, Cheng-PRL01} and the International Space Station.\cite{BaileyPRL07,ManleyPRL2004} The main findings, with a focus on solidification and phase separation, will be briefly reviewed in the following. Although there certainly are important effects of gravity on the formation of colloidal particles as well (see e.g. Vanderhoff \textit{et al.}\cite{Vanderhoff1986}), these will not be considered here. A few studies have been performed during the last 15 years to elucidate these effects, going from Earth milligravity to Space microgravity experiments. The Chaikin and Russel groups performed experiments in the Space Shuttle Columbia to study the effects of gravity on the crystallization of colloidal hard spheres.\cite{Zhu-Nature97, Cheng-PRL01} Microgravity experiments performed in the coexistence region ($49.4\% \leq \phi \leq 54.5\%$) showed a face-centred cubic (f.c.c.) structure, suggesting that f.c.c. is the equilibrium structure for colloidal hard spheres.\cite{Cheng-PRL01} This was confirmed by computer simulations.\cite{WoodcockNature97,FrankelNature97} On the other hand, colloidal crystals grown just above the melting volume fraction ($\phi \geq 54.5\%$) exhibited a random stacking of hexagonally close-packed planes (r.h.c.p.), whereas in normal gravity they showed a mixture of r.h.c.p. and f.c.c. packings, suggesting that the f.c.c. component was formed as a result of gravity-induced stresses.\cite{Zhu-Nature97} In the same experiments in space, dendritic growth of colloidal crystals was observed. This kind of crystal growth was inhibited in normal gravity, because the rough dendrites were simply sheared off by the flow caused by the sedimentation of the colloidal crystals. At much higher volume fractions, glassy systems, which failed to crystallize on Earth, crystallized fully in less than two weeks in microgravity. Simeonova \textit{et al.}\cite{Simeonova04} used real-space Fluorescent Recovery After Photobleaching (FRAP) to study the influence of gravity on the long-time behavior of the mean-square displacements (MSD) of glassy polydisperse colloidal hard spheres. Using two density matching conditions, corresponding to different gravitational lengths: $l_{\mathrm g}$ of 0.2 mm (low gravity effect) and 13 $\mu$m (normal gravity), the authors showed that under the effect of normal gravity and for samples prepared at volume fractions higher than $\phi > 55\%$, aging takes place over a much wider time window. That explains the observation that colloidal hard sphere systems which are glassy on Earth rapidly crystallize in Space.

Other experiments were performed to study the effects of gravity on colloidal dispersions. Bailey \textit{et al.}\cite{BaileyPRL07} used small-angle light scattering and direct imaging to study the phase separation in a deeply quenched colloid-polymer mixture in microgravity on the International Space Station (ISS). The authors followed this process for nearly 5 decades in time, from the early stage of spinodal decomposition, which is governed by diffusion-limited dynamics, to the late stage governed by interfacial-tension-driven coarsening, where the domains reached a centimeter in size. The phase separation dynamics closely mimic that of molecular fluid mixtures after very shallow quenches through the critical point. Manley \textit{et al.}\cite{ManleyPRL2004} studied the aggregation kinetics of polystyrene colloids in Space and compared the results with those measured on Earth. In microgravity the clusters grew at a rate in qualitative agreement with that expected for diffusion-limited cluster aggregation (DLCA), which persisted for the full duration of the Space experiment. By contrast, the time evolution of the average cluster sizes on Earth showed a significant deviation from DLCA behavior. The initial growth of the clusters was identical to that in Space, following a power-law time evolution consistent with DLCA; however, after a certain time, the growth deviated from the expected behavior and reached a plateau, with no further growth observed. Okubo \textit{et al.}\cite{OkuboCPS2000} studied the kinetics of colloidal alloy crystallization in binary mixtures of polystyrene and silica particles having different sizes and densities in microgravity using parabolic flights in an airplane. The growth rates increased substantially in microgravity up to about 1.7 times those in normal gravity, which was mainly caused by the absence of particle segregation that took place under normal gravity. Tsuchida \textit{et al.}\cite{TsuchidaCPS2000} studied the effect of added salt on the crystallization kinetics of silica particles. The authors found that the rate coefficients decreased as the salt concentration increased and that those in microgravity were smaller than those in normal gravity. Dokou \textit{et al.}\cite{Dokou} used Atomic Force Microscopy to study the effect of gravity on the deposition of polystyrene sulfate latex, silica and colloidal gold particles onto solid surfaces. The results showed that significant differences are observed in the particle deposition onto horizontal and vertical surfaces, under identical suspension conditions and exposure times. When the density of the colloidal particles is high, as in the case of colloidal gold particles, gravity can become a significant driving force for particle transport to the surface.

As mentioned, only a limited number of experiments can be and have been performed in true microgravity conditions. It is therefore important to diagnose whether gravity may significantly influence the outcome of a ground-based experiment. Bartlett \textit{et al.}\cite{Bartlett91,Bartlett94} proposed a simple and elegant way of simulating time-averaged \textit{zero}-gravity conditions on earth. The method consists simply of rotating the samples slowly (with ca. 1 Hz) in the vertical plane at a certain angular velocity, $\omega$, so the sedimentation of a particle is effectively averaged out over one cycle. The optimum angular frequency is determined by a balance between the rate of sedimentation and the magnitude of the rotationally induced shear field within a sample. Time scales of other dynamical processes such as phase separation should be taken into account as well. In fact, $\omega$ must be sufficiently large to ensure that composition gradients at any instant of time are not significant over the diameter of for instance an aggregate. Furthermore, $\omega$ must also be sufficiently small to ensure that the rotationally induced shear gradient and/or convection does not significantly distort the equilibrium microstructure. With our experiments we roughly aimed at reducing settling of particles within one rotation to the particle size or less. With this technique, the authors studied binary mixtures of large and small spherical particles, and compared the results with those subjected to normal gravity. The overall volume fraction of the suspension, $\phi_{\mathrm{tot}}=\phi_A + \phi_B$ was 56.6\% with a mixing ratio of AB$_{32}$. Under normal gravity, in the top 1 mm of the system an irregularly stacked close-packed crystal of small spheres (species B) formed together with an ordered AB$_{13}$ alloy structure. Lower down in the sample, the authors observed 15 mm of a single crystalline phase of small spheres, with the large spheres in amorphous structure. At the bottom, the system formed 10 mm of an amorphous structure. The observation of AB$_{13}$ confirmed the importance of gravity in these binary mixtures since equilibrium studies had shown that AB$_{13}$ structures are formed only in suspensions with compositions between AB$_{10}$ and AB$_{16}$.\cite{SchofieldPRE05} By contrast, after slowly rotating the sample for 138 days and standing ($\omega = 0$) for 40 hours, the dispersions remained homogeneous and crystallization proceeded throughout the sample with a crystalline phase of small spheres with the remaining, relatively small number of, large spheres in amorphous structures.

Amazingly, the methodology of Bartlett \textit{et al}. has not seen that much follow-up work, although it was reinvented at least once in work on the effects of depletion on micron-sized colloidal platelets by Badaire \textit{et al}.\cite{BadaireJACS2007} In this paper, we will, mostly qualitatively, study several systems that we feel are representative of many self-assembling systems being studied by many groups. We investigate the difference between systems left to evolve under the influence of gravity and the same systems for which a slow continuous rotation averaged out the effect of particle sedimentation (see Fig .\ref{rotstage}). Ultimately, it is our goal to find out under which conditions the methodology of averaging out gravity will give similar results as true \textit{zero}-gravity experiments and under which conditions it will not. Although it is not part of the present paper, having a confocal scan head attached to the rotation stage is in principle possible, and may provide single particle level trajectories of the systems that in this paper are studied more qualitatively. The systems we studied include: a hard sphere fluid, a colloidal crystal of particles interacting via long-range repulsions, an oppositely charged colloidal gel that is crystallizing, a colloidal system with short-range depletion attraction and long-range electrostatic repulsion, dipolar colloidal chains, and colloidal gold platelets with attractive van der Waals interactions. To do so, we prepared identical pairs of samples of these different systems. For each case, the sample was rigorously homogenized on a vortex apparatus and transferred into two glass capillaries. Subsequently, one sample was stored horizontally under normal gravity (see Fig. \ref{rotstage}.c, except in the case of charged colloidal crystal (CCC) where the sample was stored verticlally and observed with a tilted confocal microscope), whereas the other one was rotated in a home-built rotating stage to average out the effect of particle sedimentation (see Fig. \ref{rotstage}.a,b,d). After a number of days, we compared the two samples by studying them with confocal microscopy. 

\section{Rotational averaging-out gravitational sedimentation}
\label{RAOGSsection}

\begin{table*}[t]
\begin{center}
\begin{tabular}{|l|l|l|l|l|}
\hline
System & Particle  & Debye  & Density    & Gravitational \\
       & diameter  & length & difference & length $l_{g}$ ($\mu$m)\\
       & $\sigma$ ($\mu$m) & $\kappa^{-1}$ ($\mu$m) & $\Delta \rho$ (g/ml) & (and P\'{e}clet number)\\
\hline
 HSF  (PMMA/CHB/TBAB salt)          & 2.2           & ca. 0.   &  ca. 0.15  & ca. 0.5 ($Pe \approx$ 2.2)\\
 CCC (PMMA/CHB)                     & 1.4           &  3 & 0.15  & 2 ($Pe \approx$ 0.35)\\
 OCCG (PMMA/CHB/decalin/TBAB)       & 1.06 and 0.91 &  0.25& 0.02 & 40 ($Pe \approx$ 0.012)\\
 CDA (PMMA/CHB/decalin)             & 0.892 &  1.6 & 0.02 & 40 ($Pe \approx$ 0.012)\\
 DCC  (Silica/DMSO)                 & 1.45          &  0.2 & 1.23 & 2 ($Pe \approx$ 0.36)\\
 CGP (Gold platelets/H$_{2}$O)           & 1.2             & 0.01 & 18.3 & 0.025 ($Pe \approx$ 24)\\
\hline
\end{tabular}
\caption{Some characteristics of the suspensions used in this study: particle diameter, Debye screening length, density difference between the particles and the solvent, gravitational length, and the P\'{e}clet number. The colloidal gold platelets (CGP) were 20 nm thick. The abbreviations used here stand for: HSF for Hard Sphere Fluid, CCC for Charged Colloidal Crystal, OCCG for Oppositely Charged Colloidal Gel, CDA for Colloids with Depletion Attraction, DCC for Dipolar Colloidal Chains, and CGP for Colloidal Gold Platelets. }
\label{TablePMMAs}
\end{center}
\end{table*}

\begin{figure}[!ht]
\center
\includegraphics[width=0.47\textwidth]{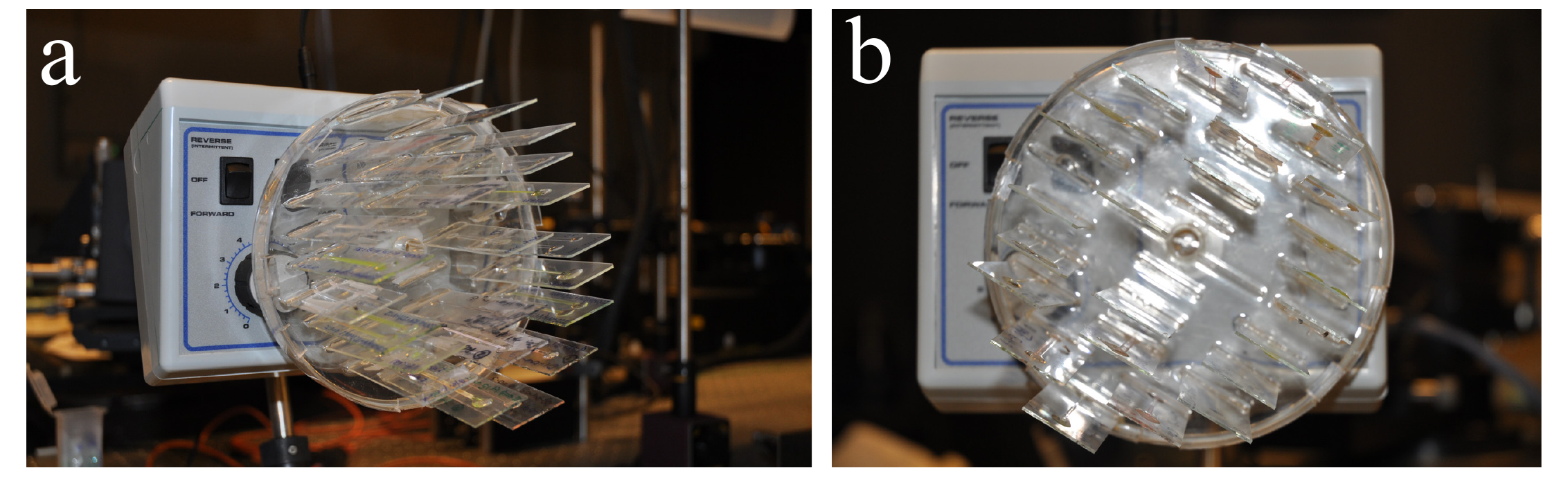}
\includegraphics[width=0.47\textwidth]{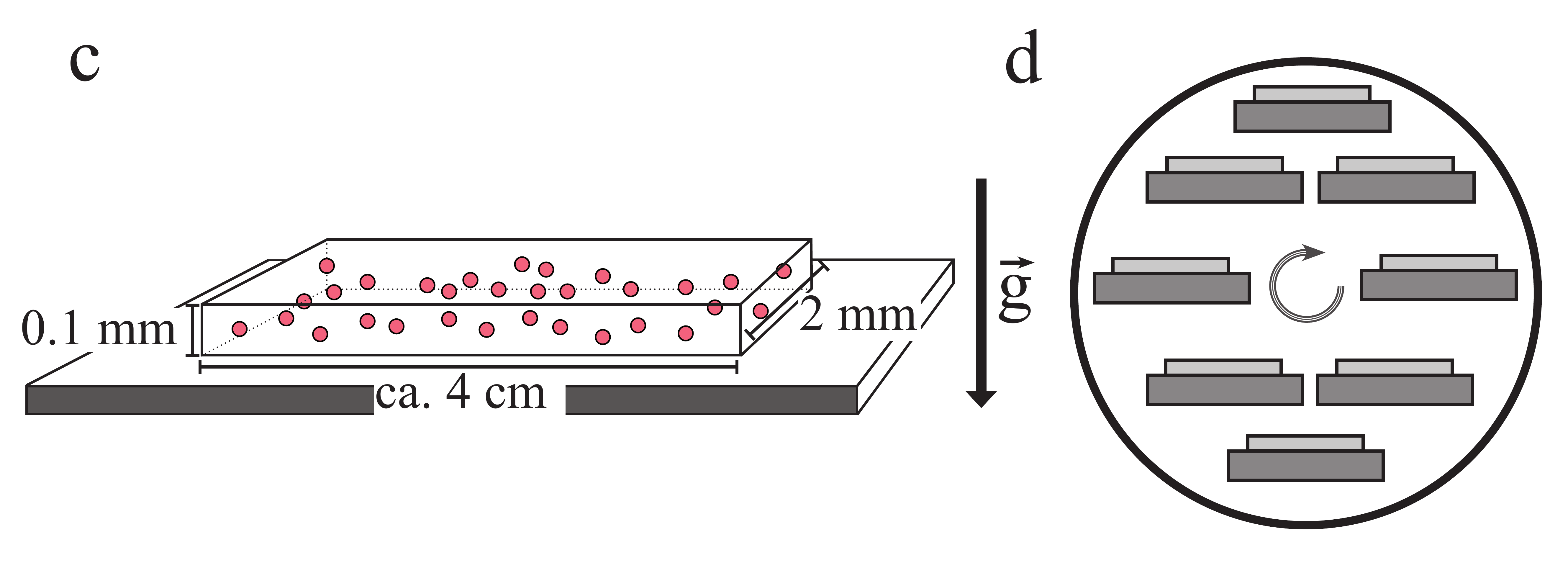}
\caption{(a,b) Pictures of the rotating stage loaded with samples on microscope slides. (c,d) Sketch of the experimental setup. Images are not to scale. (c) Sketch of a sample consisting of glass slice with a glass capillary mounted onto it and stored horizontally. The capillary contains the colloidal dispersion. (d) Samples mounted on a stage that is slowly rotating in a plane oriented parallel to the direction of gravity. The length direction of the capillary is pointing out of the paper. The optimum angular frequency is determined by a balance between the rate of sedimentation and the magnitude of the rotationally induced shear field within a sample. As a result, the sedimentation of colloids is effectively time-averaged out. We rotated samples at a rate of 3 rpm in this work.}
\label{rotstage}
\end{figure}

To average out the effects of gravitational sedimentation, we built a rotating stage using the design by Badaire \textit{et al.}\cite{BadaireJACS2007} (see Fig.~\ref{rotstage}.a,b) that enabled us to rotate samples at rates between 0 and 10 rpm. This design consists of a sample holder, made from a silicone elastomer, in which several slots were molded to accommodate microscope glass slides, and a peristaltic pump installed in such a way that the sample holder attached to its rotor rotates in a vertical plane. The fabrication of the stage required 4 main steps: 1) a mold was fabricated from a rigid assembly of 28 vertically and uniformly distributed glass slides (26 mm long, 1 mm thick) inserted into the lid of a 145 mm plastic Petri dish. The vertical position of the slides was adjusted such that a gap of approximately 1 mm remained at the bottom of the Petri dish after positioning the lid over it. 2) A silicone elastomer was prepared by mixing 100 mL of Sylgard DC 184 elastomer base and 10 mL of curing agent. Once degassed in a dessicator connected to a vacuum pump, the elastomer was poured into the Petri dish, and the lid with the glass slide assembly was placed over it. 3) The whole assembly was placed in an oven at 80$^{\circ}$C for 24 h to cure the elastomer. 4) Once at ambient temperature, the sample holder was separated from the mold with the help of a few drops of ethanol, and screwed to the rotor of an inexpensive peristaltic pump (variable-flow peristaltic pump, Fisher Scientific), installed on a support in such a way that the sample holder rotates exactly in a vertical plane parallel to gravity. 

We also constructed a setup in which it was possible to rotate the samples while simultaneously putting them under the influence of a well defined electric field. The setup was build from an inexpensive motor and gear assembly and a mercury wetted rotating contact (Mercotac  Model 305). The mercury in this "slip ring" ensures there is absolutely no contact bounce present in the signal. With this setup an electric signal of 250V with a frequency ranging from DC to 1 MHz could be applied to the rotating sample. Since the samples in the stage rotate slowly, sedimentation is effectively averaged out and additionally, convection is reduced. However, since the confocal microscope setup is presently not designed for rotation, particle sedimentation could not be averaged out during the observation of the samples. We believe that the observed structures were not affected by gravity during the short time ($1-2$ min) between taking the samples out of the rotation stage and the 3D-imaging. 

The different dispersions studied in this work were transferred each into two glass capillaries of $0.1 \times 2.0$ mm inner dimensions and approximately $4$ cm long (VitroCom, see Fig. \ref{rotstage}.c). The samples were stored horizontally under the influence of gravity (except in the case of charged colloidal crystal (CCC) where the sample was stored verticlally) and on the rotating stage for several days. The capillaries were treated in an identical way as well and were closed with the same glue (Norland optical adhesive), which was cured for the same amount of time with UV light ($\lambda = 365$ nm, UVGL-58 UV lamp, UVP). Care was taken to avoid inclusion of an air bubble in the sample, which could disrupt colloidal structures when on the rotating stage. With the confocal microscope, we observed that in both capillaries for each system the same state was present initially. Table \ref{TablePMMAs} gives some details about the different systems used in this study, labeled HSF for the Hard Sphere Fluid, CCC for Charged Colloidal Crystal, OCCG for Oppositely Charged Colloidal Gel, CDA for Colloids with Depletion Attraction, DCC for Dipolar Colloidal Chains and CGP for Colloidal Gold Platelets.

\section{Results and discussion}

\subsection{Hard Sphere Fluid (HSF)}

\textit{Hard-sphere} colloidal systems are one of the simplest models used to study a variety of fundamental questions. The phase diagram of these systems is well known, \cite{Pusey86} showing a fluid phase at a volume fraction less than $49.4\%$, a fluid/crystal coexistence at $49.4\% \leq \phi \leq 54.5\%$, and between $54.5\%-74\%$ a close packed crystal. When crystallization is avoided by introducing particle size polydispersity, hard sphere suspensions become very viscous and eventually form an amorphous solid at large volume fraction. \cite{PuseyPRL89}

\begin{figure}[!ht]
\center
\includegraphics[width=0.31\textwidth]{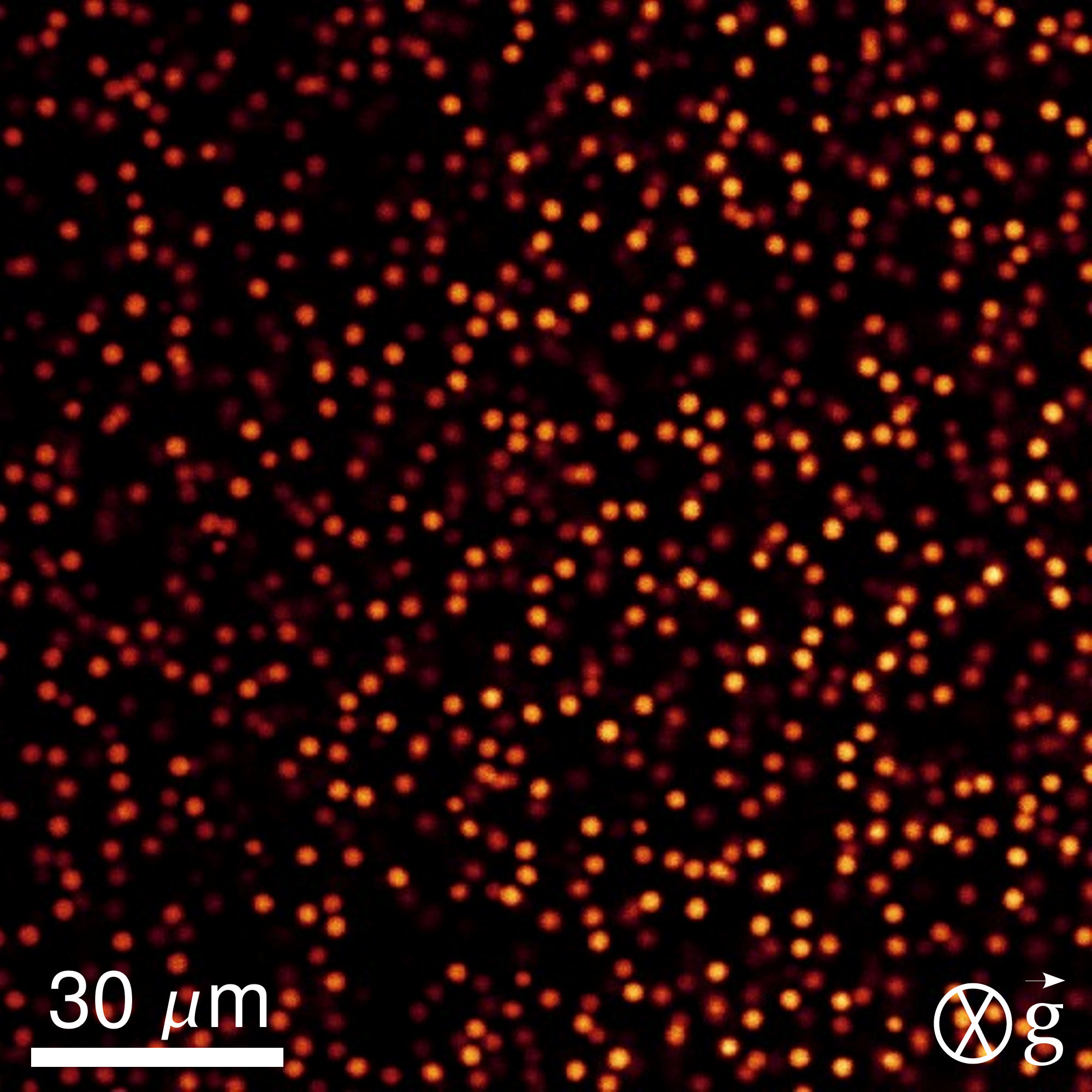}
\caption{Confocal image of a hard sphere fluid (volume fraction $\phi \sim 10 \%$) made of PMMA particles, suspended in cyclohexyl bromide (CHB) saturated with 260 $\mu$M salt tetrabutyl ammonium bromide (TBAB).}
\label{Fluid5percPMMACHB}
\end{figure}

In this section, we investigate the effects of gravity on a dilute fluid made of fluorescently labelled and sterically stabilized poly methyl-methacrylate (PMMA) particles with a diameter of 2.2 $\mu$m and size polydispersity of $3\%$, dispersed in cyclohexyl bromide (no density matching, $\Delta \rho \approx 0.15$ g/ml) saturated with added salt tetrabutyl ammonium bromide (260 $\mu$M TBAB, Sigma-Aldrich). Note that the particles sediment upwards (creaming) since they have a lower density than the solvent. In these experiments on sterically stabilized and refractive index matched colloids,\cite{Pusey86,YethirajNature2003} the interparticle interactions can easily be tuned from long-range electrostatic repulsive to (almost) hard sphere-like, by adding salt that allows to effectively screen the colloidal surface charge.\cite{YethirajNature2003,RoyallPRL07,Royall-CondMatt03} The suspension was prepared at a volume fraction $\phi \sim 10\%$, and was transfered into two glass capillaries of $0.1 \times 2.0$ mm inner dimensions (VitroCom). The samples were stored horizontally in one case and on the rotating stage in the other case. With the confocal microscope, we observed that in both capillaries the same fluid-like state (see Fig.~\ref{Fluid5percPMMACHB}) was present initially. 

\begin{figure}[!ht]
\center
\includegraphics[width=0.45\textwidth]{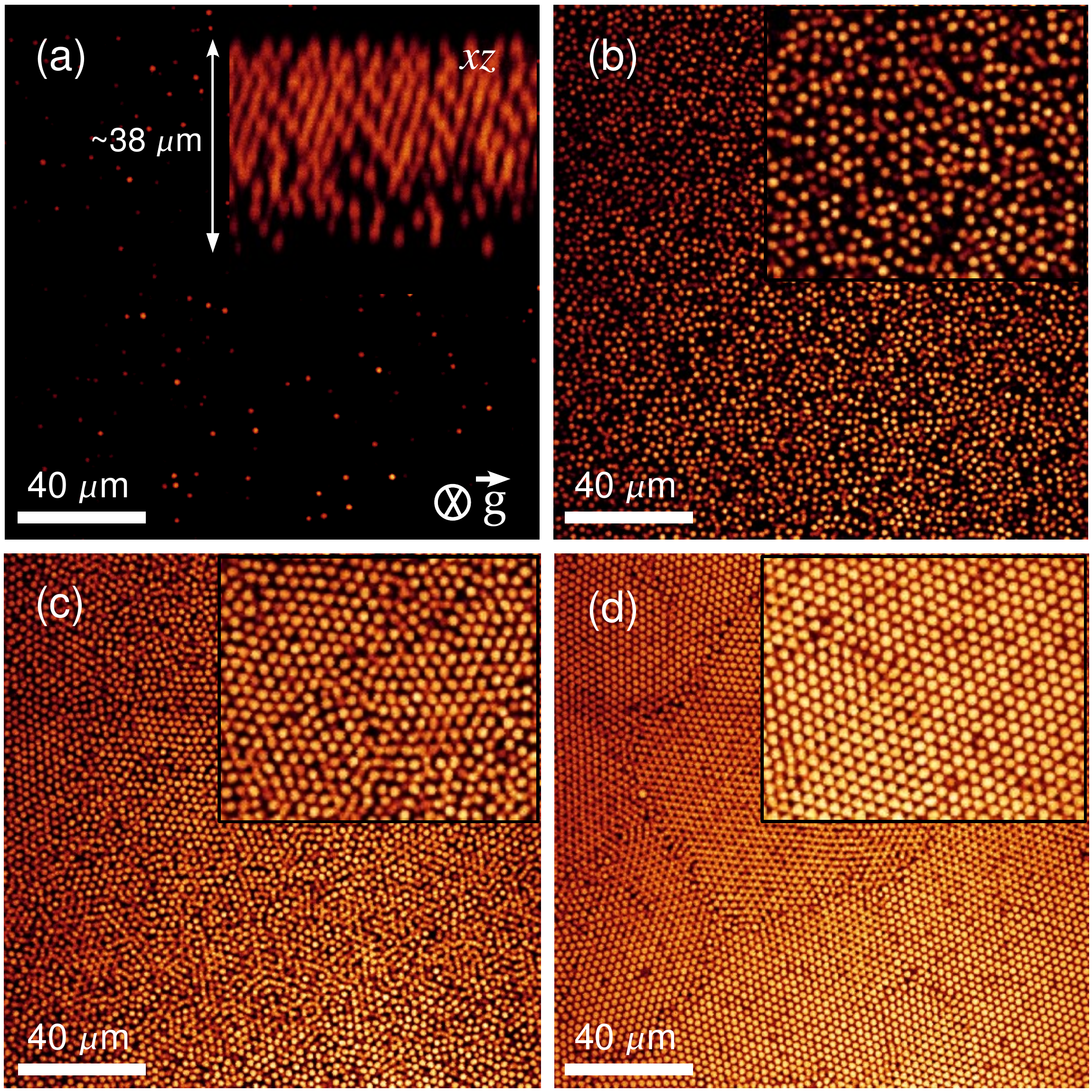}
\caption{Confocal images of the sedimented fluid after the equilibrium density gradient had established (note that the particles sediment upwards since they have a lower density than the solvent): (a) very dilute fluid ($\phi\lesssim 2\%$) at a height $z\simeq 32 ~\mu$m, (b) dense fluid phase ($\phi \lesssim 40\%$) at $z\simeq 25 ~\mu$m, (c) fluid/crystal coexistence ($\phi \lesssim 53\%$) at $z\simeq 18 ~\mu$m and (d) close packed crystal ($\phi \lesssim 70\%$) at $z\simeq 6 ~\mu$m. All distances are measured from the top of the cell. The inset in (a) shows an $xz$ image of 38 $\mu$m sediment, and in (b), (c) and (d) show 2 times zoomed images.}
\label{SedimentedFluid5percPMMACHB}
\end{figure}

After a few tens of minutes, significant sedimentation was already observed in the sample that was stored horizontally. Two time scales are relevant: that of diffusion; here the Brownian time is defined as the time it takes a particle to diffuse over a distance of one radius, $\tau_B=3\pi \eta \sigma^3/4 k_B T \approx 15$ seconds, and that of gravity induced settling, which can be characterized by the time it takes a particle to sediment over one radius, $\tau_S=9 \eta/\Delta \rho g \sigma \approx 7$ seconds. As already mentioned in the introduction, the ratio of these time scales is quantified by the P\'{e}clet number $Pe=\tau_B/\tau_S \approx 2$ (gravitational length $l_g\approx 0.5 ~\mu$m). Royall \textit{et al}.\cite{RoyallPRL07} and Schmidt \textit{et al}.\cite{SchmidtJPCM08} quantitatively studied the nonequilibrium sedimentation of hard-sphere dispersions with P\'{e}clet number $Pe \sim 1$, that were confined in horizontal capillaries using confocal microscopy, dynamical density functional theory (DDFT), and Brownian dynamics simulations. The authors showed that DDFT gives accurate results as compared to computer simulations and that it describes laterally homogeneous sedimentation very well as compared to experimental data. Hoogenboom \textit{et al.}\cite{HoogenboomJCM02,HoogenboomJCM03} studied in real-space crystallization and the occurrence of stacking faults formed through sedimentation of (soft) silica colloids with P\'{e}clet numbers on the order of $10^{-1}$. The crystals showed a high amount of f.c.c-stacking compared to bulk crystallization of hard-spheres. In our case, the system has a stacking probability not close to that of f.c.c, but closer to r.h.c.p (random hexagonal closed packed structure). Note that the density profile could have been measured and from this the equation of state could have been obtained.\cite{BeckhamJCP07} 

After a few hours, crystallization occured on the top of the capillary cell (PMMA particles cream in CHB). Note that the time needed for a particle to sediment over the capillary thickness (ca. $100~\mu$m) is on the order of a few tens of minutes. On the other hand, particles in the sample stored in the rotating stage were homogeneously distributed and did not display any significant density gradient. Observed one week later, the particles in the sample that was stored horizontally reached an equilibrium density profile (Fig.~\ref{SedimentedFluid5percPMMACHB}). However, no significant change was observed on the rotated sample. We have checked that rotating the hard sphere fluid does not alter the particle distribution, by comparing the measured radial distribution functions (not shown here) for the sample just after preparation and after more than a week of a storage in the rotating stage. 

\subsection{Charged Colloidal Crystal at low ionic strength (CCC)}

\begin{figure}[!ht]
\center
\includegraphics[width=0.31\textwidth]{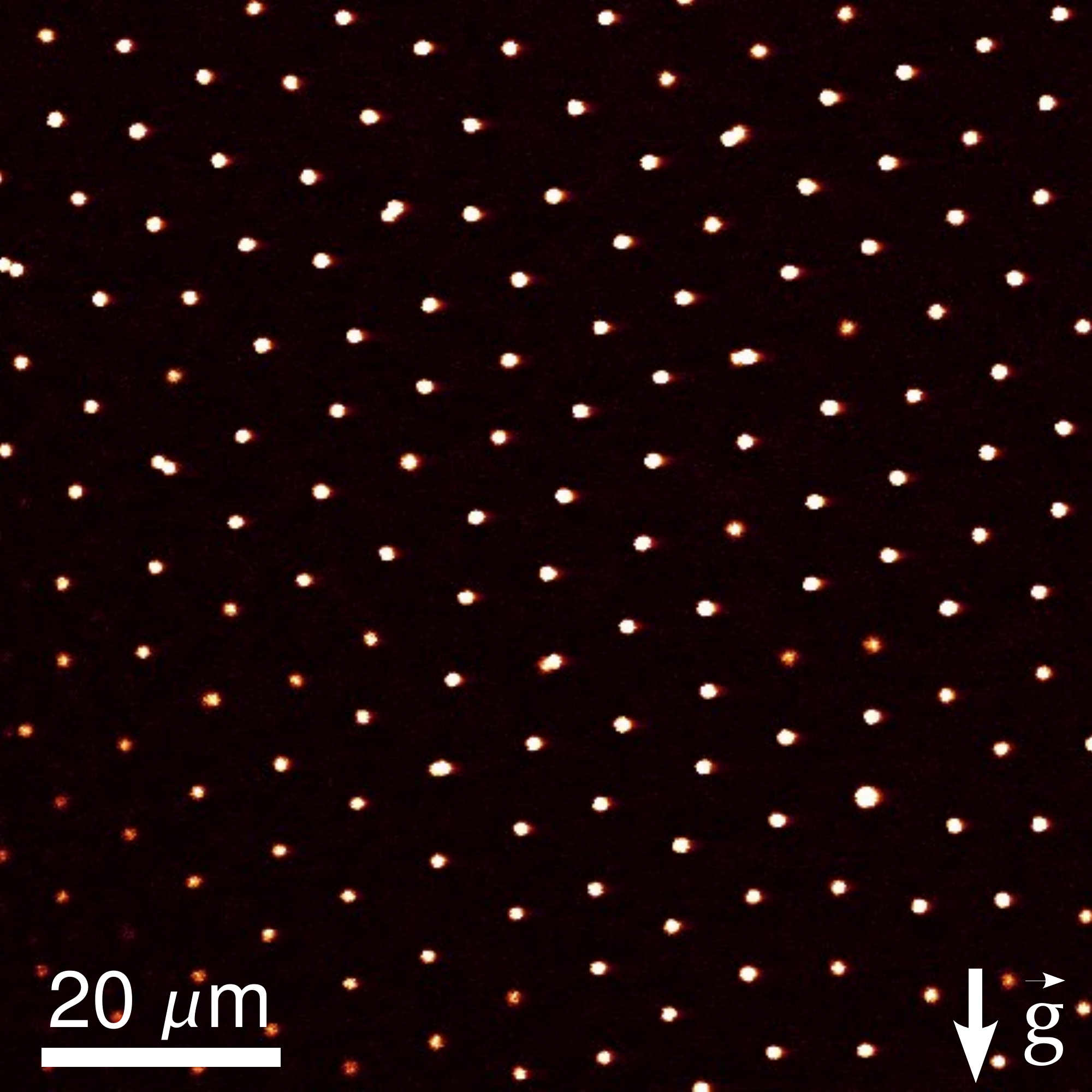}
\caption{Confocal microscopy image of a (110) plane of a long-range body-centred cubic (b.c.c) crystal of sterically stabilized poly methyl-methacrylate particles (PMMA) with diameter of $1.4~\mu$m, suspended in pure cyclohexyl bromide (CHB). The volume fraction $\phi \sim 1\%$ and the interparticle spacing is $d \approx 10 ~\mu$m.}
\label{Crystal1percPMMACHB}
\end{figure}

Colloidal particles, suspended in a solvent with a (relative) dielectric constant roughly between 4 and 15, can form very long-range crystal lattices even for micron-sized particles when the
suspension is carefully purified to remove free ions from the suspension. \cite{Royall-CondMatt03,YethirajNature2003,MirjamPCCP07} A suspension of PMMA particles suspended in purified cyclohexyl bromide CHB (dielectric constant $\epsilon = 7.6$) is one example of such systems. In these kinds of solvents, charge dissociation still occurs spontaneously \cite{VanderHoeven92}, contrary to truly apolar media ($\epsilon \approx 2$) that require charge stabilizing surfactants.\cite{HsuLangmuir05} The particles used in this study carried a positive charge, $Z > 500$ electrons (as quantified by means of electrophoresis measurements of the dilute system), and were surrounded by a double layer of compensating negative charges. Each particle exerts a significant screened Coulomb force on its neighbors, confining the Brownian motion to excursions around lattice positions because the array crystallizes under the influence of purely repulsive forces. The interparticle spacing can be as much as ten times the particle radius, as shown in Fig.~\ref{Crystal1percPMMACHB}, which is typical of \textit{Wigner crystals}.\cite{Wigner,LeunissenPNAS07,IrvineNature10} 

We observed the effect of gravitational compression on a long ranged crystal of PMMA particles with diameter of $1.4 ~\mu$m, suspended in pure cyclohexyl bromide (CHB) at a volume fraction $\phi \sim 1\%$ (see Fig.~\ref{Crystal1percPMMACHB}). To reach a very low ionic strength condition, we purified the solvent CHB using a procedure described elsewhere,\cite{Pangborn} and put the suspension in contact with water that serves as a reservoir absorbing the free ions from the suspensions.\cite{MirjamPCCP07} The conductivity of CHB as received (ca. $7000$ pS/cm) decreased to less than $20$ pS/cm with the purification procedure. Due to the small volume of the dispersion contained inside the capillaries, we could not directly measure the conductivity of the oil-water combination in our experiment, but it was previously shown\cite{MirjamThesis} that upon contact with an equal volume of water the conductivity of CHB decreases by a factor of 10. From this, we estimated the Debye screening length to be $\kappa^{-1} \sim 3~\mu$m in this case. The crystal structure observed at this low particle volume fraction and for our experimental condition of the screening $\kappa \sigma \sim 0.5$ has a b.c.c symmetry, as b.c.c under these conditions has a lower free energy state than f.c.c.\cite{HynninenPRE03, RoyallJCP06} The capillary cell was placed vertically in this case and observed with a 90$^{\circ}$ tilted confocal scanhead combined with an accurate translation stage that allows to obtain 3D data stacks at all the different heights along the length of the capillary (which is ca. 3 cm in this case). The water phase was located above the CHB phase, since water has a lower density than CHB. 

\begin{figure}[!ht]
\center
\includegraphics[width=0.46\textwidth]{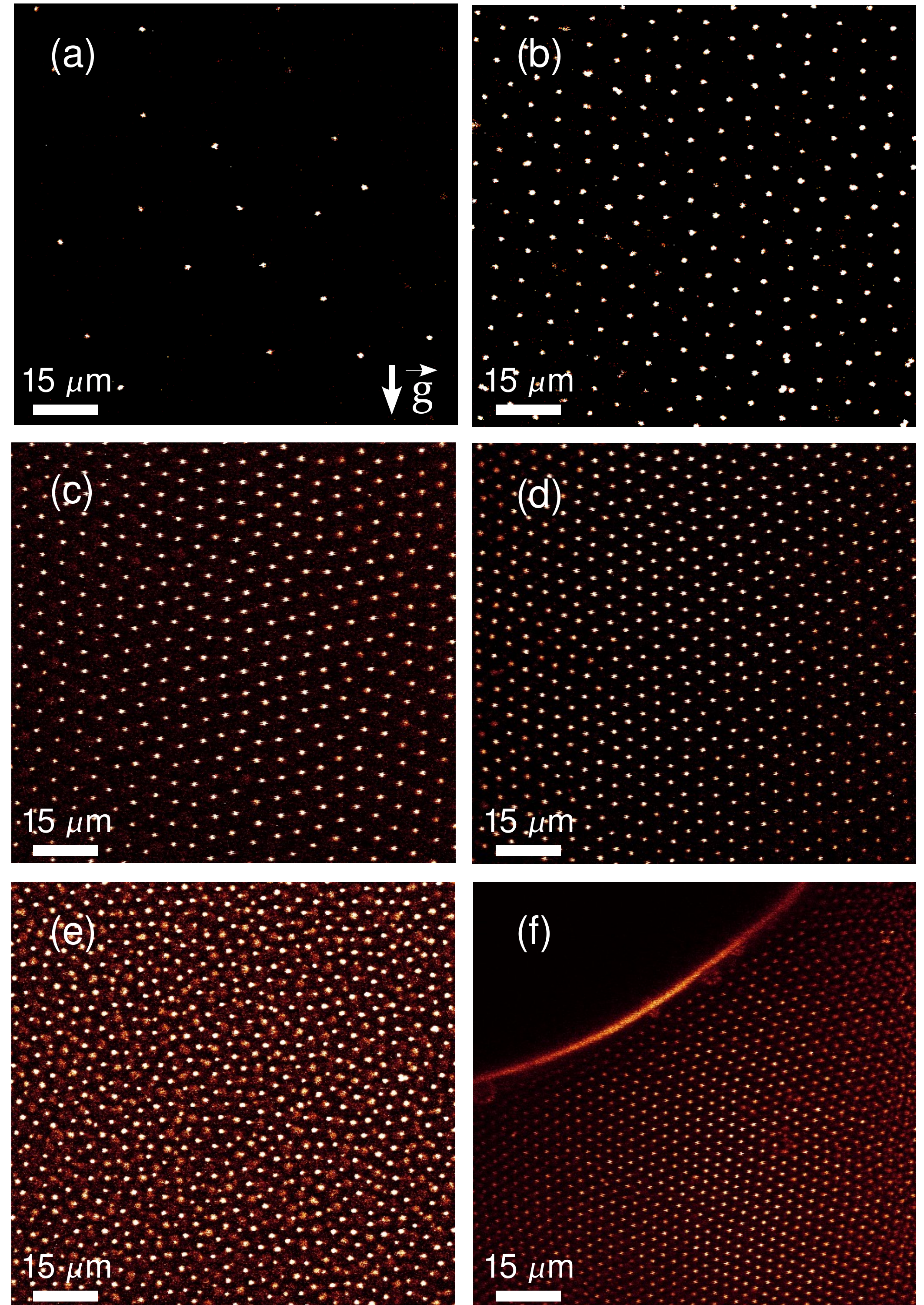}
\caption{A nonequilibrium situation of an initially homogeneous phase of a b.c.c crystal of PMMA particles suspended in pure CHB after the sample was stored vertically more than 1 day under the influence of gravity, showing a sequence of dilute-Fluid/dilute-Crystal/dense-Fluid/dense-Crystal phases. Figure (f) shows the interface with water, that allows to continuously keep the system at a low ionic concentration. (a) Dilute fluid ($\phi < 0.2\%$) at a height $z=28$ mm below the water interface. Long range crystals: (b) a (110) plane of a b.c.c crystal ($\phi \simeq 1\%$) at $z=20$ mm, (c) a (110) plane of a b.c.c crystal ($\phi \simeq 5\%$) at $z=16$ mm and (d) a (111) plane of a f.c.c crystal ($\phi \simeq 9\%$) at $z=12$ mm. The lattice spacing varies in the crystal phase from more than 10 $\mu$m to 3-4 $\mu$m. (e) A reentrant dense fluid ($\phi \simeq 10\%$) at $z=8$ mm (particles diffuse freely) and (f) a (111) plane of a f.c.c crystal ($\phi > 10\%$).}
\label{Sequence}
\end{figure}

Figure \ref{Sequence} shows the nonequilibrium situation of the system after it was stored vertically for 1 day under the influence of gravity. The system formed a dilute fluid (in the lowest region of the capillary because the particles have a lower density than the solvent), followed by long-range crystals with different lattice spacings, varying from more than 10 $\mu$m to 3-4 $\mu$m. Going further upwards, we observed a reentrant dense fluid phase until finally the system crystallized again at a much higher volume fraction. At this low ionic strength condition and increasing the particle volume fraction by gravity, the system showed a transition from a fluid phase, to low density crystals of b.c.c structure, to f.c.c crystals, as confirmed previously by experiments and numerical simulations.\cite{HynninenPRE03,SirotaPRL89} A dramatically different phase behavior was observed in the other (initially identical) capillary that was left in the rotating stage. During all this time and throughout the whole capillary the low volume fraction crystal as depicted in Fig.~\ref{Crystal1percPMMACHB} was preserved. 

The unusual sequence dilute-Fluid/dilute-Crystal/dense-Fluid/dense-Crystal phases observed throughout the suspension under the influence of gravity is likely a consequence of the volume fraction dependence of the interaction potential, an observation made before on a similar system by some of the authors.\cite{Royall-CondMatt03} In more recent work, we measured interaction forces between pairs of charged PMMA colloidal particles suspended in an equivalent solvent (cyclohexyl chloride, CHC) directly from the deviations of particle positions inside two time-shared optical traps.\cite{DjamelSoftMatter2011} In addition, from electrophoresis experiments it was observed that the particle charge in similar systems as the one used here decreases non-linearly with increasing volume fraction.\cite{TeunJCIS11} Current work is in progress using this method to probe the interaction forces in this system more directly with optical tweezers in more concentrated situations as observed in this section, which will allow for a better understanding of several of the open questions related to, for instance, the presence of non-additive many-body interactions in these  long-ranged charged systems.\cite{ReinkePRL07,BrunnerPRL04,SianisPRL07}

After more than 3 days, the observed sequence of reentrant crystal and liquid phases collapsed into a less than 2 mm thick layer of dense crystals with few layers of a fluid phase just below. We note again that the dilute crystal on the rotating stage remained long ranged for more than a week of observation. As mentioned above, we believe that this intriguing behavior observed in this system while it is in transition to its equilibrium state in the gravitational field is caused mainly by a density-dependent surface charge in our system. Particles lose their charges during densification but not completely, which explains why the system does not reach the hard sphere crystal. At this point we conjecture that this effect of colloidal discharging\cite{RoyallJCP06} is responsible for the huge variation of the lattice spacing in the sedimented crystal and for the observed reentrant fluid phase. Indeed, densification of the system under gravity and, consequently, the colloidal discharging make the competition between electrostatic interaction energy and thermal energy quite complex.\cite{RasaNature04,RoyallJPCM05} These effects are the subject of work in progress. 

\subsection{Oppositely Charged Colloidal Gel (OCCG)}

Since a few years, binary systems of oppositely charged colloids are used to create so-called \textit{ionic colloidal crystals} in low polar solvent mixtures.\cite{LeunissenNature05, BartlettPRL05, HynninenPRL06, Hynninen2PRL06} Particles can be made oppositely charged by tuning the salt concentration and the volume fractions of both particle species.\cite{TeunThesis2010} When the attractions are strong ($U_{\textbf{AB}} \gg k_B T$), the oppositely charged particles strongly attract each other and become kinetically trapped in a gel-like structure.\cite{SanzJPCB08,SanzJPCM08} When the charges are sufficiently low and the interactions are on the order of $1~k_B T < U_{\textbf{AB}} < 10~k_B T$, the system can efficiently explore phase space, resulting in the formation of oppositely charged binary crystals. The phase diagram for such systems was calculated, and CsCl and CuAu and random f.c.c structures were found both in computer simulations and experiments.\cite{LeunissenNature05,BartlettPRL05,HynninenPRL06,Hynninen2PRL06} 

\begin{figure}[!ht]
\center
\includegraphics[width=0.33\textwidth]{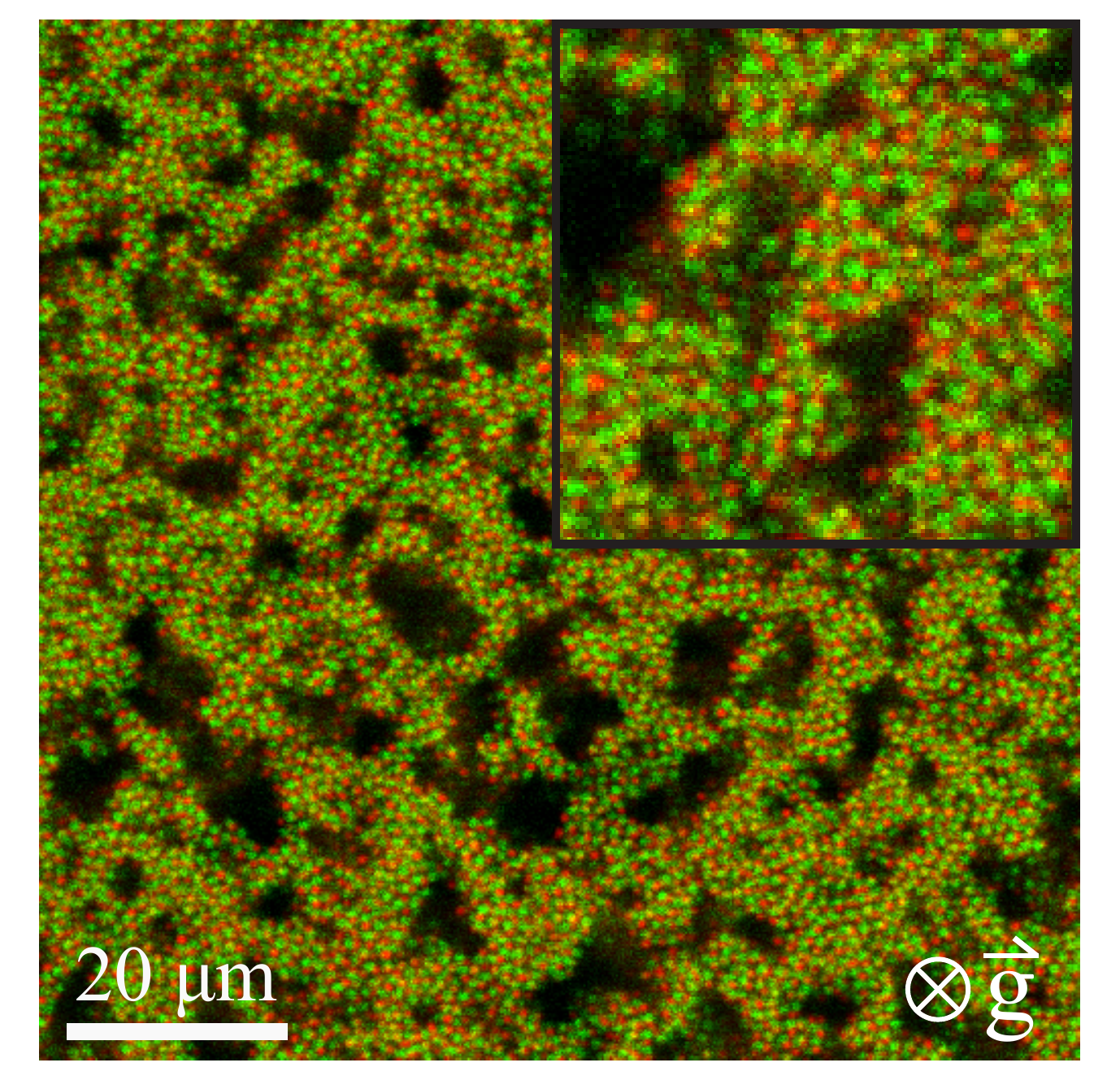}
\caption{A gel-like structure formed after transferring a mixture of oppositely charged colloids, suspended in CHB/cis-decalin with $26$ $\mu$M TBAB, to a capillary cell. The total volume fraction $\phi_{\mathrm{tot}} = 30\%$.}
\label{ch5:day0small}
\end{figure}

To study the gravity induced collapse of a gel-like structure made of oppositely charged colloids, we prepared a 1:1 mixture (in volume) of PMMA particles with diameters of 1.06 and 0.91 $\mu$m (see table \ref{TablePMMAs}) in a nearly, but not perfectly, density matched mixture of  CHB/cis-decalin (27.2 w\%) containing tetrabutyl ammonium bromide salt (26 $\mu$M TBAB, Sigma-Aldrich) (see Fig. \ref{ch5:day0small}). The estimated Debye screening length is $\kappa^{-1} \approx 0.25~\mu$m and the suspension was prepared at a total volume fraction $\phi_{\mathrm{tot}} = 30\%$. 

\begin{figure}[!ht]
\center
\includegraphics[width=0.46\textwidth]{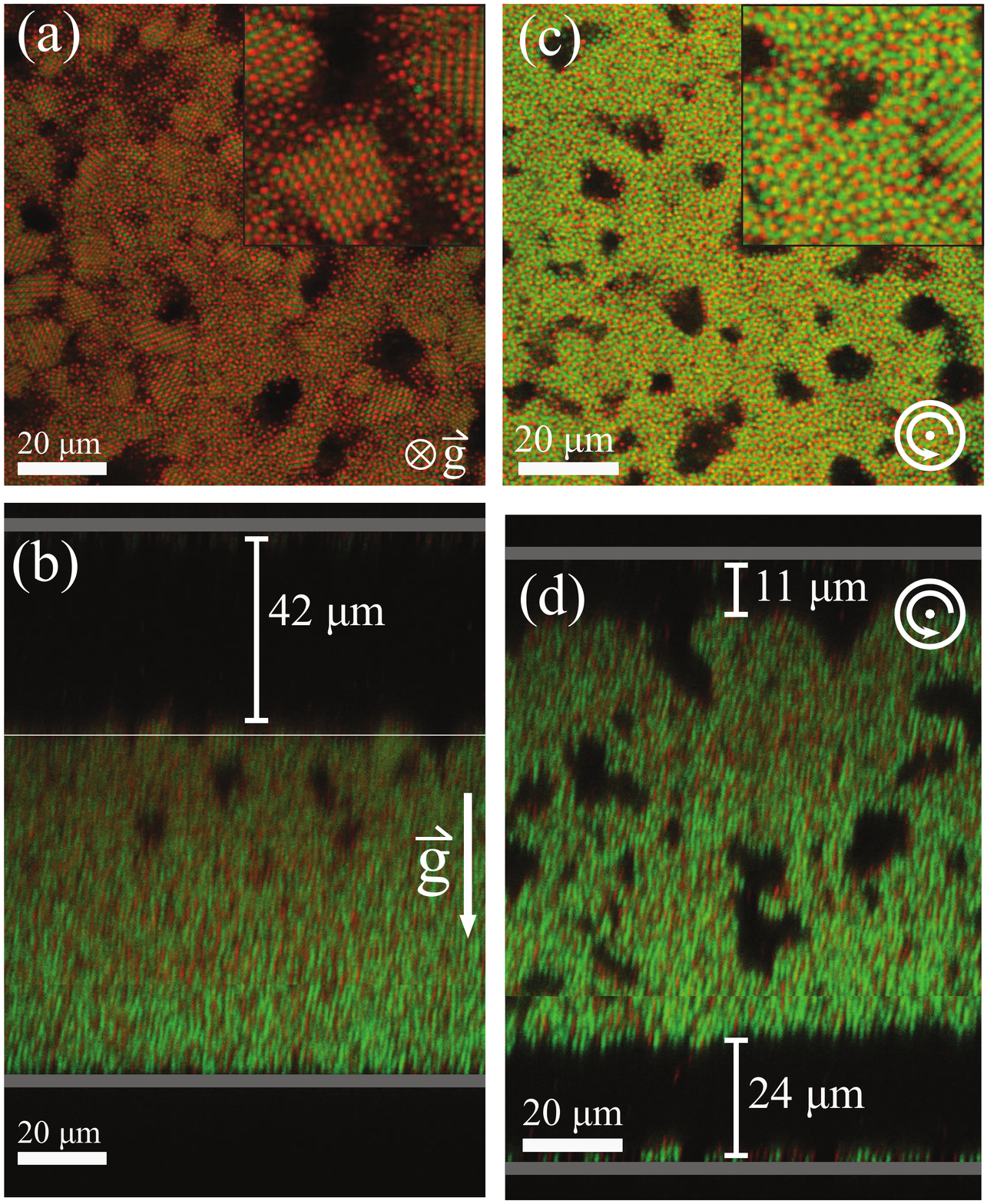}
\caption{The suspensions after $18$ days of horizontal storage (a,b) or in the rotating stage (c,d). Due to sedimentation, the gel-like structure had partially collapsed and CsCl crystal domains (ca. $5-30$ $\mu$m in size) were present on top of the sediment (grey line in (b)). The distance from the lower glass wall to the top of the sedimented part (including the crystals) was $75$ $\mu$m. For the sample stored in the rotating stage, the gel-structure survived and small crystal domains (ca. $5-15$ $\mu$m in size) had emerged around the voids inside it. The thick grey horizontal lines in (b,d) denote the position of the capillary glass walls. Image (b) and (d) are composite images that are build up from two or three confocal images.}
\label{ch5:day18small}
\end{figure}

After $18$ days of horizontal storage, the gel had partially collapsed and the particles were visibly sedimented (Fig.~\ref{ch5:day18small}b). Crystal domains of only several layers thick were present on top of the sediment (Fig.~\ref{ch5:day18small}a). For the sample that was stored on the rotating stage, the gel-structure was still present and no signs of sedimentation were observed (Fig.~\ref{ch5:day18small}d). The gel had contracted and pulled away from both walls due to the cohesive forces. Interestingly, crystal formation occurred at the interface between the gel and the voids inside the gel (Fig.~\ref{ch5:day18small}c). 

\begin{figure}[!ht]
\center
\includegraphics[width=0.47\textwidth]{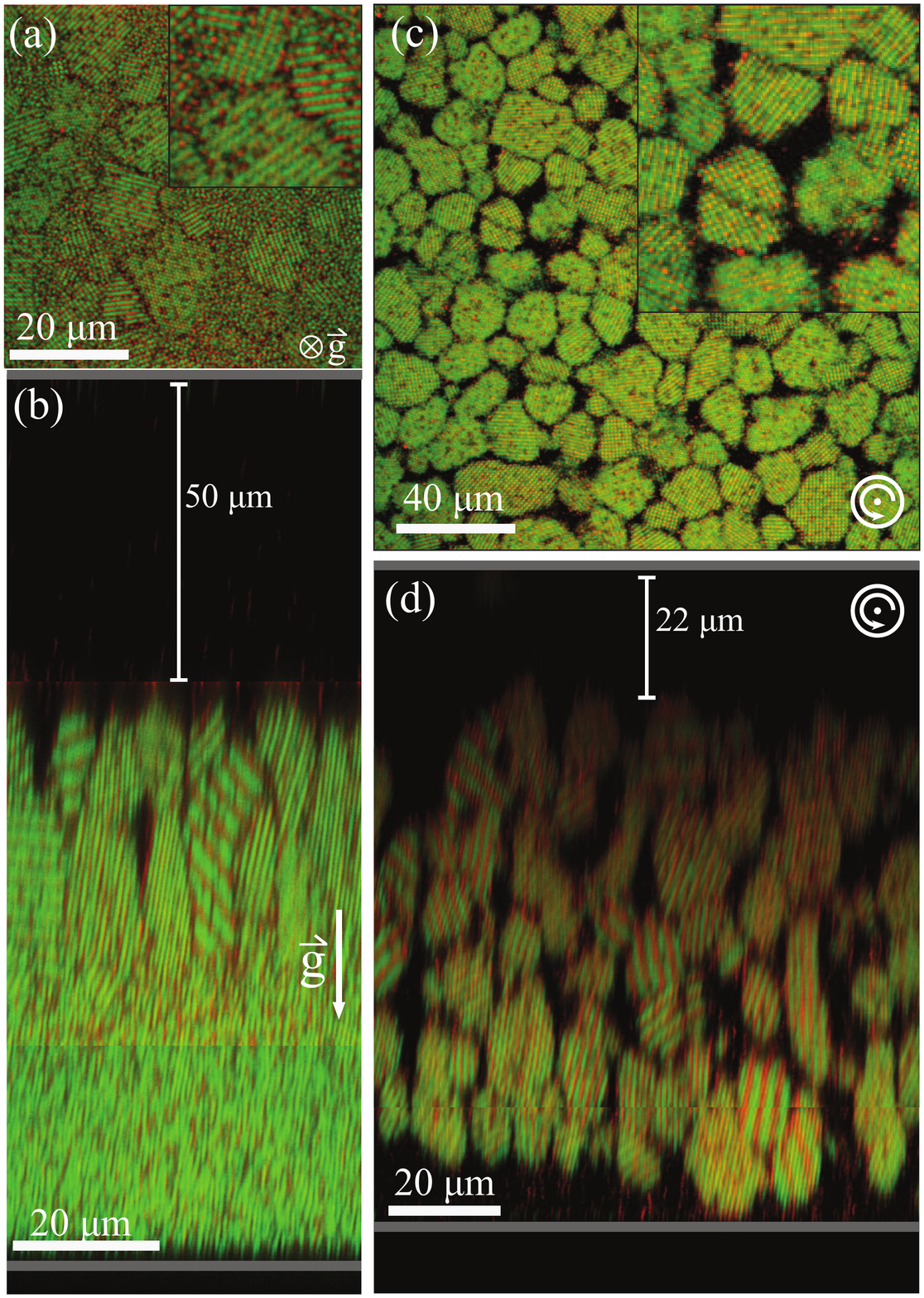}
\caption{The suspensions after $30$ days of horizontal storage (a,b) or in the rotating stage (c,d). Under the influence of gravity, the crystals on top of the sediments that were already present after $18$ days had grown longer in the direction parallel to the gravitational field. The distance from the lower glass wall to the top of the crystals was ca. $75$ $\mu$m. On the rotating stage, the gel-like structure had completely disappeared and all particles were present in crystal domains. There were large voids between different crystal domains. The grey horizontal lines in (b,d) denote the positions of the capillary glass walls. Image (b) and (d) are composite images that are build up from two or three confocal images.}
\label{ch5:day30small}
\end{figure}

After spending a total time of $30$ days of horizontal storage, the crystals that were present at the top of the sediment had grown out in an elongated fashion (Fig.~\ref{ch5:day30small}b). Since the distance from the top of the crystallites to the lower glass wall of the sample did not change much compared to the situation after $18$ days (Fig.~\ref{ch5:day18small}a,b), the crystal growth direction must have been downwards into the collapsed gel. Apparently, the voids in the sediment offered sufficient translational freedom for particle reorganization that was required for the conversion of the sediment into the crystal. Approximately $40$ $\mu$m above the lower glass wall, the growth process stalled, suggesting that the sediment became too dense to allow for the particles to rearrange. Contrary to this, for the sample that was stored for a total time of $30$ days on the rotating stage, the gel-like structure had completely disappeared and crystallites had appeared with random orientations throughout the whole sample (Fig.~\ref{ch5:day30small}c,d). The orientation of the crystal domains appeared random, indicating that the small crystals that were already present after $18$ days had continued to grow until all of the gel had converted into crystal. The large voids that were present between different crystals show that colloidal sedimentation was efficiently averaged out. 

\begin{figure}[!ht]
\center
\includegraphics[width=0.465\textwidth]{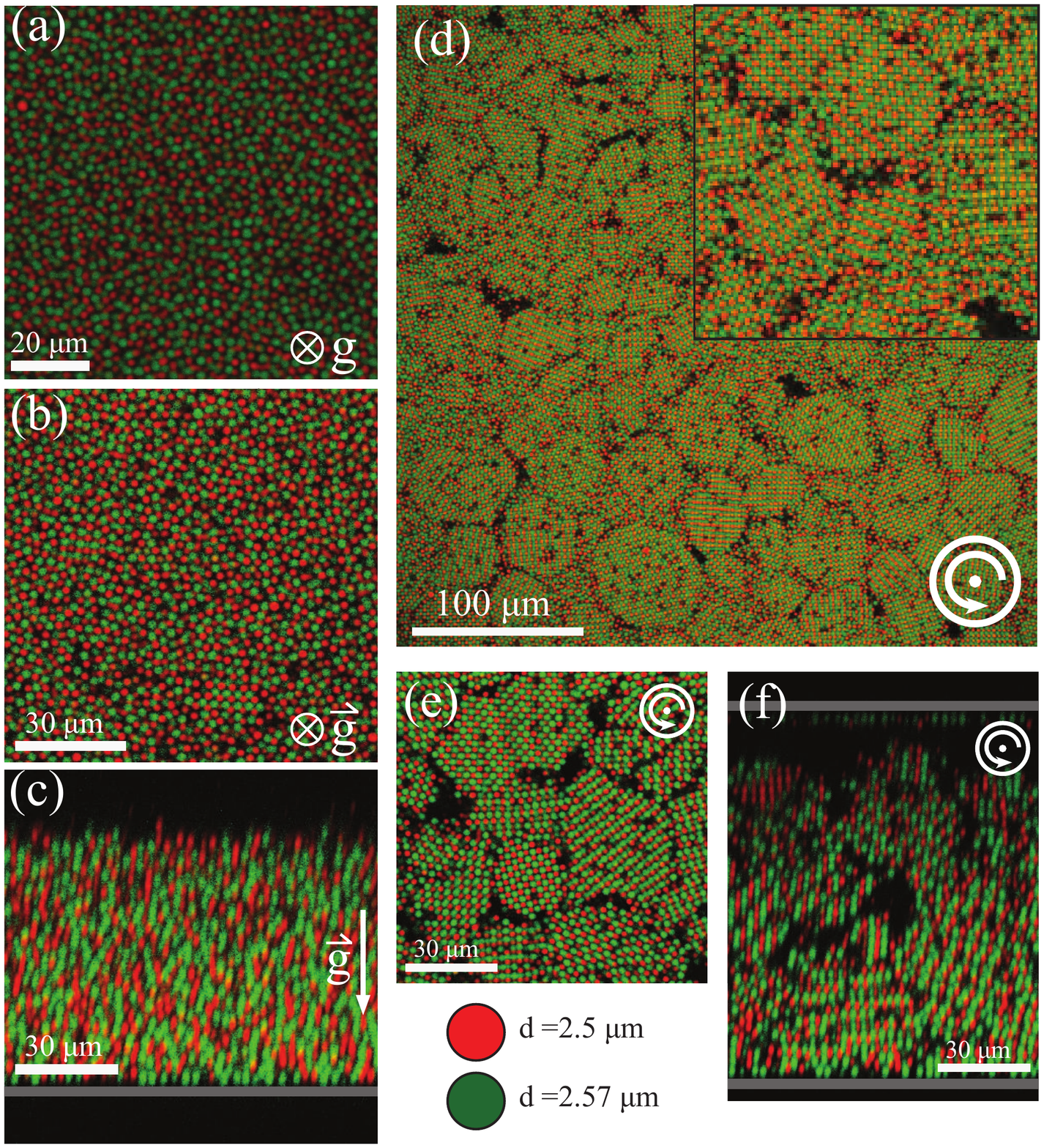}
\caption{ (a) Fluid mixture of PMMA particles with diameters of ca. $2.5~\mu$m in CHB/cis-decalin with $5$ $\mu$M TBAB after vigorous shaking ($\phi_{\mathrm{tot}} = 29\%$). (b, c) Dispersions after $68$ hours of horizontal storage show that the particles sedimented. On top of the sediment, some very small crystalline domains were visible. (d-f) After $68$ hours, the dispersion on the rotating stage contained randomly oriented crystal domains with large voids in between, similar to Fig.~\ref{ch5:day30small}. The thick grey horizontal lines in (c,f) denote the positions of the capillary glass walls. The upper glass wall in (c) is not shown here.}
\label{ch5:day0big}
\end{figure}

Additionally, we examined crystal growth starting from a fluid of oppositely charged particles (with diameter of $\sigma \approx 2.5~\mu$m in this case) at a total volume fraction, $\phi_{\mathrm{tot}} = 29\%$, (see Fig.~\ref{ch5:day0big}a). Under the influence of gravity the particles sedimented within $68$ hours, reducing the region containing particles to approximately $70$ $\mu$m (Fig.~\ref{ch5:day0big}c). On top of the sediment, small crystals were sporadically present (Fig.~\ref{ch5:day0big}b). The identical dispersion that had spent the same time on the rotating stage, however, had crystallized completely (Fig.~\ref{ch5:day0big}d-f). Similar to the gel-samples, gravity and the local volume fraction play, as expected, an important role in the crystal growth of micron-sized colloids. Our results show that for these systems of oppositely charged particles, despite being nearly density matched, densification by sedimentation makes crystal growth much slower and in some cases inhibits it almost completely.

\subsection{Colloids with short-range Depletion Attraction and long-range electrostatic repulsion (CDA)}
Contrary to the previous system of oppositely charged colloids, attraction between colloidal particles can also be induced by adding non-adsorbing polymers  to the dispersion.\cite{AsakuraJCP54,AsakuraJPS58,VrijPAC76,DeHekJCIS81,PoonACIS97,LekerkerkerBook11} This polymer-mediated \textit{depletion interaction} has received significant attention in the last decades. By varying the polymer concentration and the size ratio between colloids and polymers, one can control the strength and the range of the interparticle interaction potential, respectively. Here we study the situation where depletion interactions are supplemented by and compete with a long-range electrostatic repulsion. At low colloid concentrations, experiments, computer simulations and theory have reported a stable colloidal cluster  phase.\cite{SegrePRL01,GroenewoldJPC01,GroenewoldJPCM04,StradnerNature04,SanchezJPCM05,SchooneveldJPCB09,CampbelPRL05,SedgwickJPCM04,SciortinoPRL04}.

\begin{figure}[!ht]
\center
\includegraphics[width=0.31\textwidth]{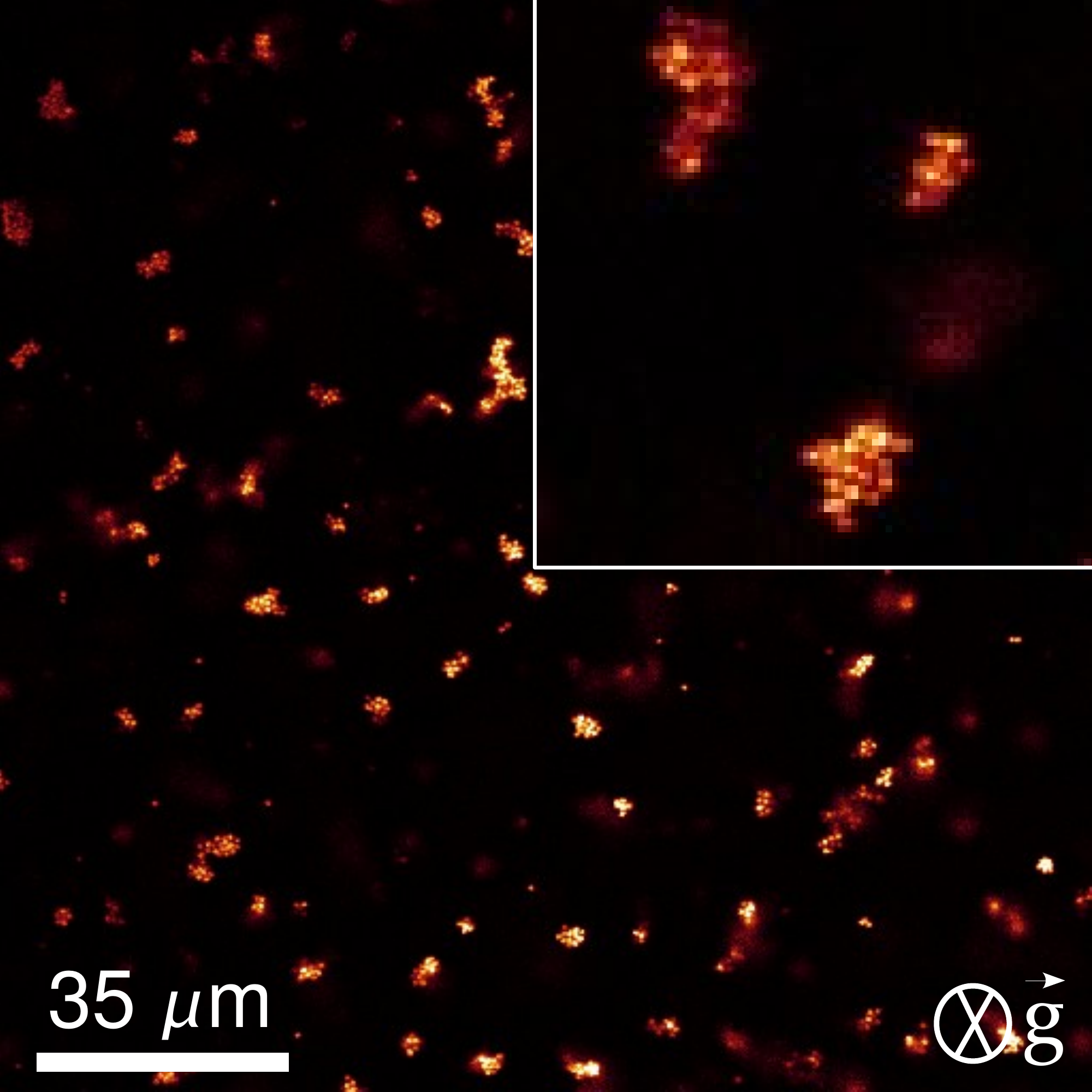}
\caption{A cluster phase of a colloidal system with competing short-range attraction and long-range repulsion, made of PMMA-particles dispersed in a mixture of cyclohexyl bromide (CHB) and cis-decalin with added non-adsorbing polystyrene polymers. The volume fraction was $\phi \sim 10\%$ and the attraction strength $U\approx-9~k_B T$.}
\label{GelDepletion}
\end{figure}

\begin{figure}[!ht]
\center
\includegraphics[width=0.465\textwidth]{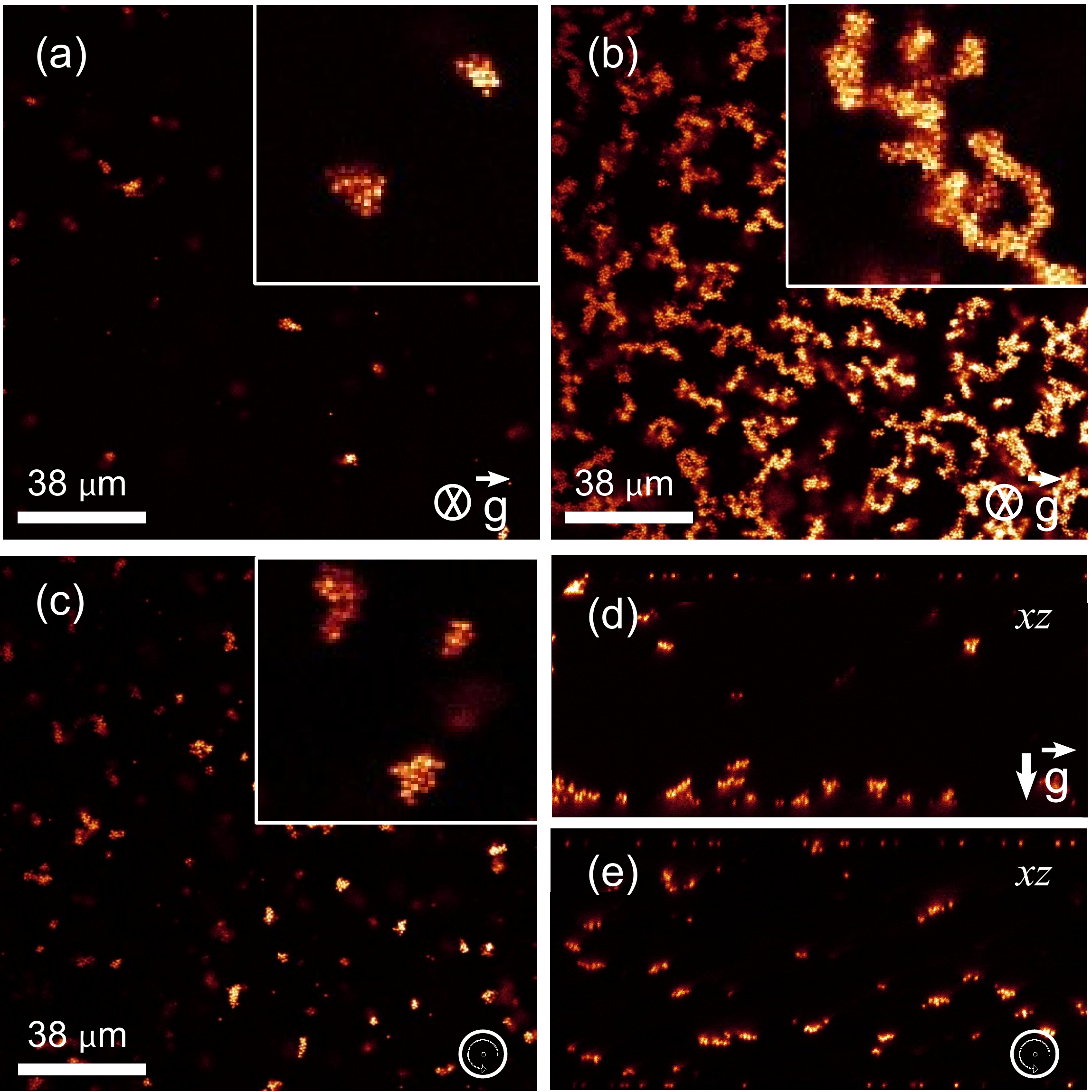}
\caption{(a-b) Confocal images of the horizontally stored sample at two different heights, $z\approx 50~\mu$m and $z\approx 7~\mu$m from the bottom of the cell, respectively. (c) Rotated sample. (d) and (e) are $xz$ images of the horizontally stored sample and the rotated one respectively. Note the absense of a gel phase in the rotated sample.}
\label{Depletion}
\end{figure}

In this section, we have verified qualitatively the effects of gravity on the cluster size distribution of a dispersion with competing short-range attraction and long-range Coulomb repulsion. In these systems, the average cluster size $R$ grows with the overall volume fraction of the colloids $\phi$ as $R \propto \phi^{1/3}$ as  predicted by theory\cite{GroenewoldJPC01} and confirmed by experiments\cite{StradnerNature04}. Sedimentation in a gravitational field couples to the local colloid volume fraction, which in turn leads to larger colloidal clusters. Ultimately these clusters aggregate amongst themselves, forming a colloidal gel. The system studied consists of PMMA-particles dispersed in a mixture of cyclohexyl bromide (CHB) and cis-decalin with a short-range depletion attraction that is induced by adding non-adsorbing polystyrene polymers. More details of this system can be found in a recent paper by Zhang \textit{et al}.\cite{ZhangPCCP09} The volume fraction of the dispersion we studied was $\phi \sim 10\%$ and the attraction strength was $U\approx-9~k_B T$. The system formed a cluster phase with clusters containing each few tens of particles on average, as shown in Fig.~\ref{GelDepletion}. We prepared two samples, one left horizontally and the other one stored in the rotating stage. After a few hours of sedimentation, the clusters had grown and became gradually interconnected, forming a gel with increasing local volume fraction (see Fig.~\ref{Depletion}b). On the other hand, the rotated sample preserved the same cluster phase during this time as depicted in Fig.~\ref{Depletion}c. Rotational averaging out gravitational sedimentation avoids gel formation in systems with short-range attraction and long-range repulsion. Instead, the cluster phase is stable. Clearly the cluster phase was not destabilized (shear-melted) by rotation of the sample. 

\subsection{Dipolar Colloidal Chains (DCC)}
Colloidal particles whose dielectric constant is different from that of the solvent, acquire a dipole moment parallel to an external electric field. The induced \textit{dipolar interactions} between the particles are anisotropic and long-ranged. The dipolar strength depends on the local particle concentration, particle size, the dielectric constant mismatch, and the applied electric field strength.\cite{YethirajNature2003,HynninenPRL2005,LeunissenAdvMatt2009,ErbNature2009} The structure of the dispersion can be tuned from strings (1D), to sheets (2D), and eventually to several different types of equilibrium 3D crystal structures by varying the dipolar field strength and concentration. In the low-field regime, the stable structure found in both experiments, theory and computer simulations is a string fluid phase that consists of chains of particles parallel to the field direction and with liquid like order perpendicular to the field direction.\cite{YethirajNature2003,HynninenPRL2005,HynninenPRE2005,DassanayakeJCP05} 

\begin{figure}[!ht]
\center
\includegraphics[width=0.31\textwidth]{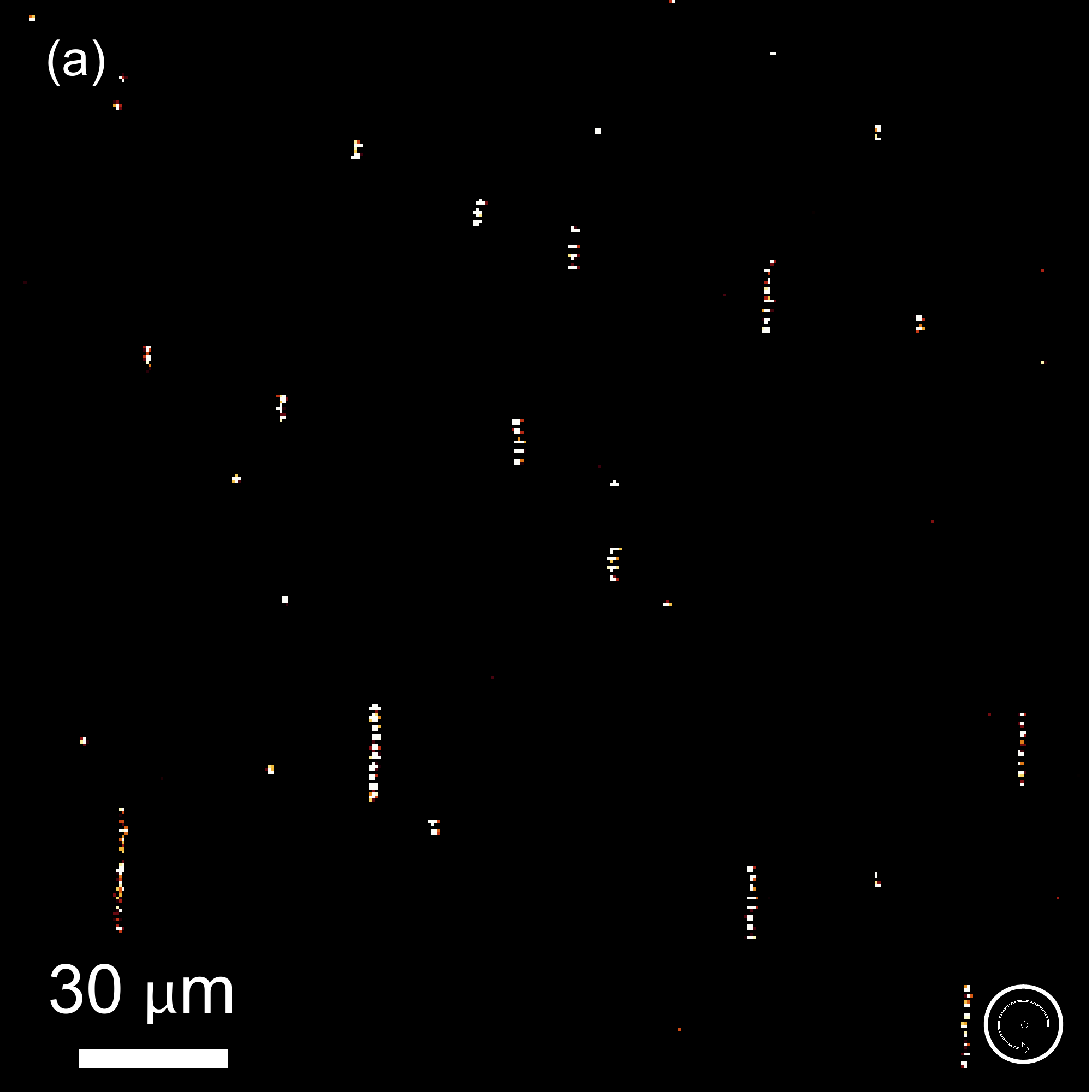}
\includegraphics[width=0.31\textwidth]{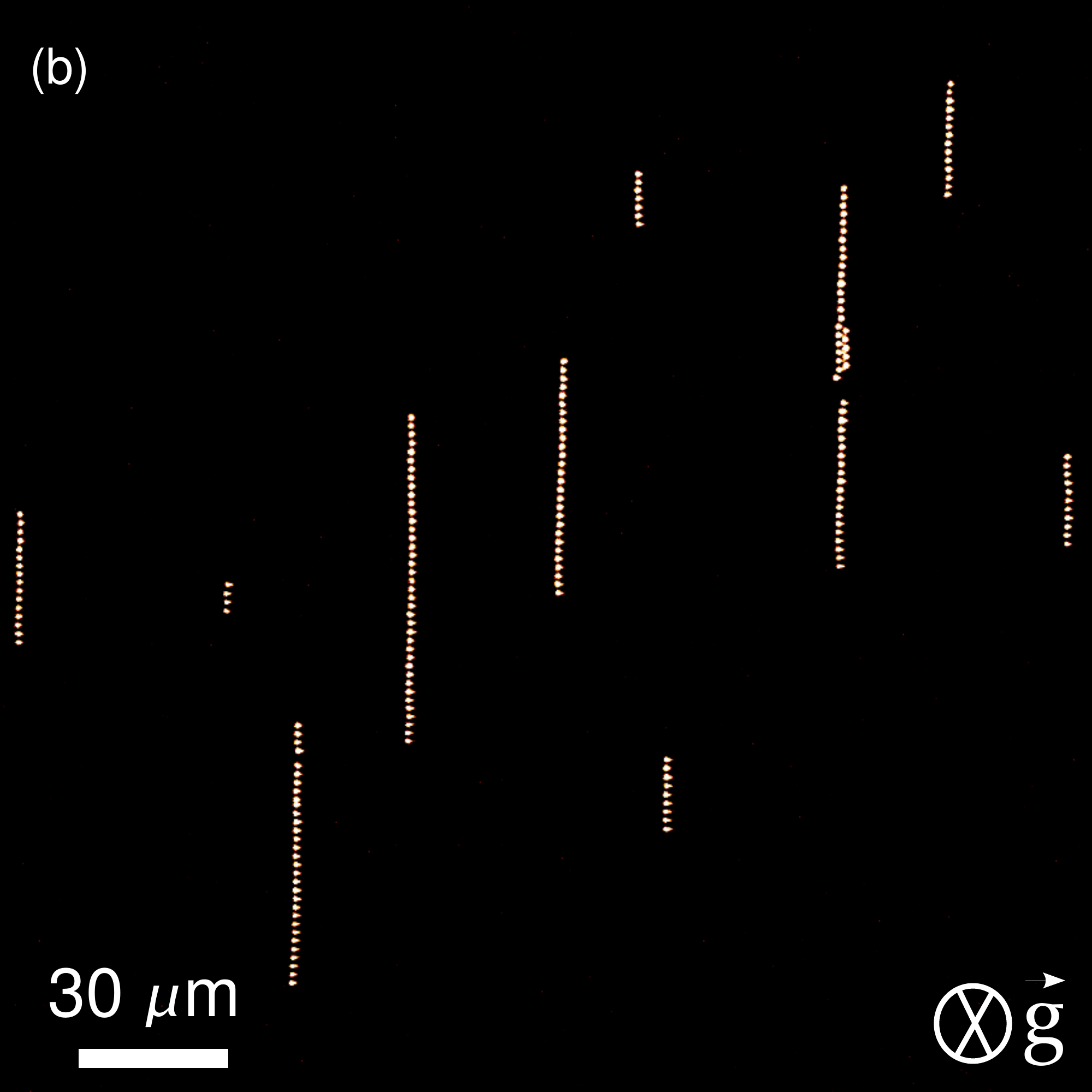}
\caption{ Confocal micrographs of silica particles in an AC electric field.  After 2 hours of waiting time, string length distribution in the two identical sample cells that were subjected to the same electric field strength: a) rotated in the vertical plane, and b) Stored horizontally.}
\label{Strings}
\end{figure}

Here, we studied the effect of gravity on the string length distribution with a system of electrostatically stabilized silica particles ($\phi < 0.1\%$, diameter $\sigma = 1.45$ $\mu$m) suspended in dimethyl sulfoxide ($\kappa^{-1} \approx 200$ nm and $\Delta \varepsilon \approx 44$) and subjected to an AC electric field. Prior to the experiment, two identical electric cells were filled with the suspension and then rotated slowly in a plane parallel to gravity to maintain the particles homogeneously dispersed throughout the cell. After 3 hours of rotation, one cell was removed from the rotating stage and then stored horizontally while the other one was left in the rotating stage. Subsequently, both cells were exposed to an external electric field. Here, we used another rotating stage that especially was designed and build to accommodate an electric field while the sample was rotating. Moreover, the electric field was supplied through a mercury contact to provide an uninterrupted connection, as explained in Sec. \ref{RAOGSsection}.

Upon application of an external electric field ($E_{rms}=0.04$ V/$\mu$m, $f$ = 1 MHz the root-mean-square electric field and the frequency respectively), the induced dipole moment in each particle led the particles to assemble into strings of one particle thick that were aligned in the field direction in a head-to-tail arrangement with a broad distribution in the length of the chains.\cite{BossisPRE07,Zubarev09} These strings then sedimented when the cell was kept horizontal, due to the density mismatch between the particles (2.1 g/ml) and the solvent (1.1 g/ml). At the bottom of the cell the local concentration of the strings slowly increased. Both ends of the longer strings continuously caught particles or small strings moving around them, thus growing into longer strings (Fig.\ref{Strings}b). The distance between the individual strings was about 20-35$\sigma$. Therefore, the dipolar contribution between these strings was unimportant and we did not observe a further change in the string length distribution. On the other hand, a distribution of short strings (Fig.\ref{Strings}a) was maintained homogeneously throughout the dispersion by means of rotating the cell. Our experimental results qualitatively demonstrate that the structure evolution, e.g. string length distribution, becomes quite different when the sedimentation of colloids is averaged out by rotating the sample. 

\subsection{Colloidal Gold Platelets (CGP)}

When dispersed in water, highly anisotropic single crystalline gold platelets ($1.2~\mu$m diameter, $16$ nm thick) can self-assemble into long wormlike columns by finely tuning the interplay between van der Waals and electrostatic interactions.\cite{Hachisu73, Hachisu76} In absence of salt, and in the limit of low volume fraction ($\phi <1\%$), the particles remain isotropically dispersed, but for ionic strengths comprised roughly between 0.8 and 2 mM, the particles form dynamic columns or stacks, the length of which depends on the depth of the secondary van der Waals minimum in the pair potential and on the volume fraction. However, because of the high density mismatch between gold and water, the particles sediment quickly in a stationary capillary and the observed structures are therefore highly influenced by gravity. Since, in this system, the main consequence of gravity is an increase of the local particle concentration near the surface, the length of the columns (at constant ionic strength) is expected to be longer after the capillary is allowed to stay horizontal for a sufficient amount of time. We verified this experimentally by comparing qualitatively the column size distribution in a suspension in which the columns were kept homogeneously dispersed throughout the whole volume with the help of the rotating stage and in an identical suspension that was kept horizontal.

\begin{figure}[!ht]
\center
\includegraphics[width=0.345\textwidth]{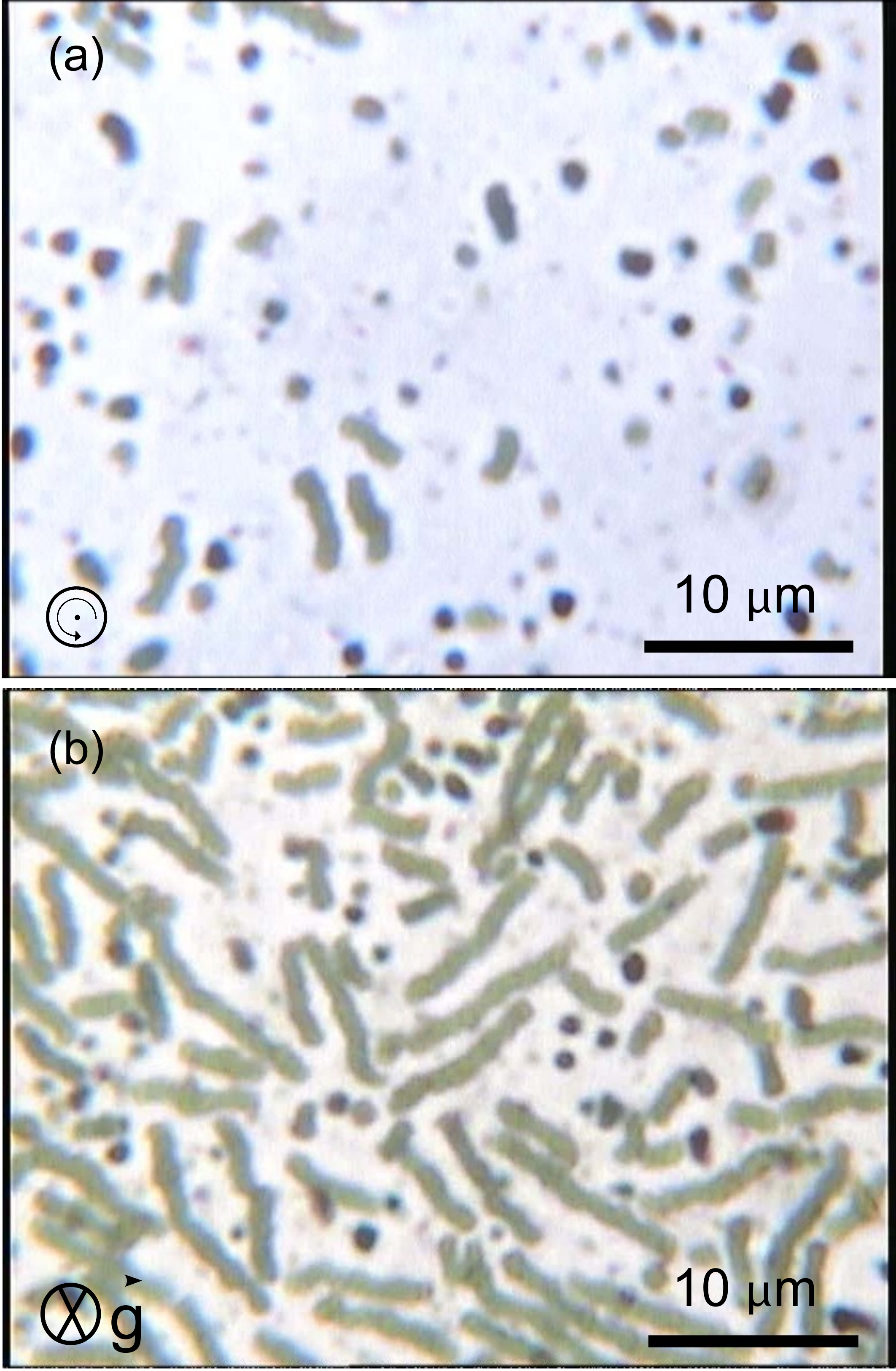}
\caption{ Highly anisotropic gold platelets assembling into wormlike columns. (a) After 72 hours on the rotating stage: short columns. (b) After 24 hours of horizontal storage: long columns.}
\label{Gold}
\end{figure} 

The first image (a) of Fig.~\ref{Gold} was captured after allowing the capillary to rotate on the stage for 72 hours after preparation (the image was taken less than 1 min after removing the capillary from the stage). The ionic strength in the suspension was 1.5 mM (NaCl). Only short columns were seen. A gas phase of very short columns and individual particles was present in the whole volume above this first layer of columns because of the very large difference in the gravitational length between individual particles and the stacks of particles. The second image (b) was captured after keeping the same capillary immobile for 24 hours. It can be clearly seen that the size distribution evolved to much larger lengths, and few short columns or isolated particles were observed. It was checked qualitatively that the same size distribution was obtained by preparing another capillary and leaving it undisturbed for 72 hours before observation. 

\section{Conclusions}
In this paper, we studied the effect of particle sedimentation on the structure formation of a whole range of different self-assembling colloidal systems, including a hard sphere fluid, a dispersion with a long Debye screening length at a density above the freezing transition, an oppositely charged colloidal gel that is crystallizing, colloids with competing short-rang depletion attraction and long-rang electrostatic repulsion, colloidal dipolar chains, and colloidal gold platelets under conditions in which they organized into equilibrium stacks. We used a simple home-built rotating stage to average out the effect of particle sedimentation, and compared the results with identical samples that were left under the influence of gravity for the same amount of time. In addition, we showed how such a setup can also be extended and used in cases were MHz frequency externally applied electric fields are used to manipulate a dispersion. In the case of the repulsive charged colloidal crystal, an unusual transient sequence of dilute-Fluid/dilute-Crystal/dense-Fluid/dense-Crystal phases was observed throughout the suspension under the influence of gravity, which is related to the volume fraction dependence of the interactions between charged particles. In the case of oppositely charged colloids, a gel-like structure was found to collapse under the effect of gravity with a few crystalline layers formed on top of the sediment, whereas the gel completely transformed into crystallites with random orientations when sedimentation was averaged out by slowly rotating the sample. We showed in general that the use of a rotating stage is very useful to diagnose when gravity may significantly influence the outcome of ground-based experiments. Moreover, we demonstrated that it can serve as a cheap and simple method to average out the effects of particle sedimentation for a wide variety of colloidal systems. Further work needs to find the conditions under which this method approaches systems at \textit{zero}-gravity preferably by comparison with similar experiments under true (micro) zero-gravity conditions. It will also be worthwhile to be able to follow structuring quantitatively in real space while the rotations are actually taking place.

\section*{Acknowledgements}
This work was supported by the research program \textit{Colloids in Space}, financed by NWO-SRON. TV and JS acknowledge NWO-CW for financial support. S.B. and H.R.V. were supported by the research program of the Stichting voor Fundamenteel Onderzoek der Materie, financed by the Nederlandse organisatie voor Wetenschappelijk Onderzoek (NWO). Part of this research was supported by the EU (NanoDirect, grant number CP-FP-213948-2).


\end{document}